\documentclass[aps,prl,superscriptaddress,twocolumn,showpacs]{revtex4}
\usepackage[pdftex,breaklinks=true,bookmarks=false,colorlinks=true,citecolor=violet,urlcolor=blue,linkcolor=red]{hyperref}
\usepackage{amsfonts,amsmath}
\usepackage{graphicx}
\usepackage{color}
\usepackage{tikz}
\usepackage{braket}

\newcommand{\comment}[1]{}

\begin{document}
 
\title{Geometric mutual information at classical critical points}

\author{Jean-Marie St\'ephan}
\affiliation{Physics Department, University of Virginia, Charlottesville, VA 22904-4714, US}
\author{Stephen Inglis}
\affiliation{Department of Physics and Astronomy, University of Waterloo, Ontario, N2L 3G1, Canada}
\author{Paul Fendley}
\affiliation{Physics Department, University of Virginia, Charlottesville, VA 22904-4714, US}
\author{Roger G. Melko}
\affiliation{Department of Physics and Astronomy, University of Waterloo, Ontario, N2L 3G1, Canada}
\affiliation{Perimeter Institute for Theoretical Physics, Waterloo, Ontario N2L 2Y5, Canada}

\date{\today}

\begin{abstract}
A practical use of the entanglement entropy in a 1d quantum system is to identify the conformal field theory describing its critical behavior. It is exactly $(c/3)\ln \ell$ for an interval of length $\ell$ in an infinite system, where $c$ is the central charge of the conformal field theory. Here we define the {\em geometric mutual information}, an analogous quantity for {\em classical} critical points. We compute this for 2d conformal field theories in an arbitrary geometry, and show in particular that for a rectangle cut into two rectangles, it is proportional to $c$. This makes it possible to extract $c$ in classical simulations, which we demonstrate for the critical Ising and 3-state Potts models.
\end{abstract}
\maketitle
 
{\bf Introduction}.---
In studies of new and exotic phases of quantum matter, the entanglement entropy
has established itself as an important resource \cite{EEreview}. It is particularly useful in the many 1d quantum critical systems governed by a conformal field theory (CFT) \cite{BPZ} in the large-distance limit. Here the R\'enyi entanglement entropy $\mathcal{S}_n$ of the ground-state is universal \cite{EE1d1,EE1d2,Korepin,EE1d3}, and the leading piece is proportional to
the central charge $c$ of the CFT  characterizing the universality class. Namely,
for a periodic system of length $L$ cut into two open segments of respective sizes $L_A$ and $L_B=(L-L_A)$, 
\begin{equation}\label{eq:hlwcc}
 \mathcal{S}_n= \frac{c}{6}\left(1+\frac{1}{n}\right)\log \left[\frac{L}{\pi}\sin \frac{\pi L_A}{L}\right]+\ldots,\ .
\end{equation}
Thus it is possible to extract the central charge from a numerical computation without fitting parameters or non-universal prefactors, and so identify the theory.
This is a striking example of the success of information-theoretic concepts applied to condensed-matter problems. 

Since a CFT also describes the large-distance limit of a two-dimensional classical critical model with rotational invariance, it is natural to expect that information-theoretic concepts can be used to analyze \emph{classical} critical systems \cite{Verstraete1,Verstraete2,Melkoinf3,Alba}. The aim of this letter is to define and compute the {\em geometric mutual information} ${\cal G}_n$, a quantity quite analogous to the quantum result in eq.~(\ref{eq:hlwcc}).  We show that in the 2d CFT case it provides an analogous quantity proportional to the central charge. For example, cutting an $L_x\times L_y$ rectangle into two $L_A\times L_y$ and $L_B\times L_y$ rectangles yields
\begin{equation}\label{eq:ad}
 {\cal G}_n=\frac{c}{2}\left(\frac{1}{n-1}\right)\log \left(\frac{f(L_A/L_x)f(L_B/L_x)}{\sqrt{L_x}f(L_y/L_x)}\right). 
\end{equation}
The function $f$ is related to the Dedekind $\eta$ function and is given in (\ref{eq:function}). As in eq.\ (\ref{eq:hlwcc}), there is no dependence on any non-universal parameters.
Below we prove this and related formulas, and give several numerical checks.

\begin{figure*}[tbp]
 \includegraphics{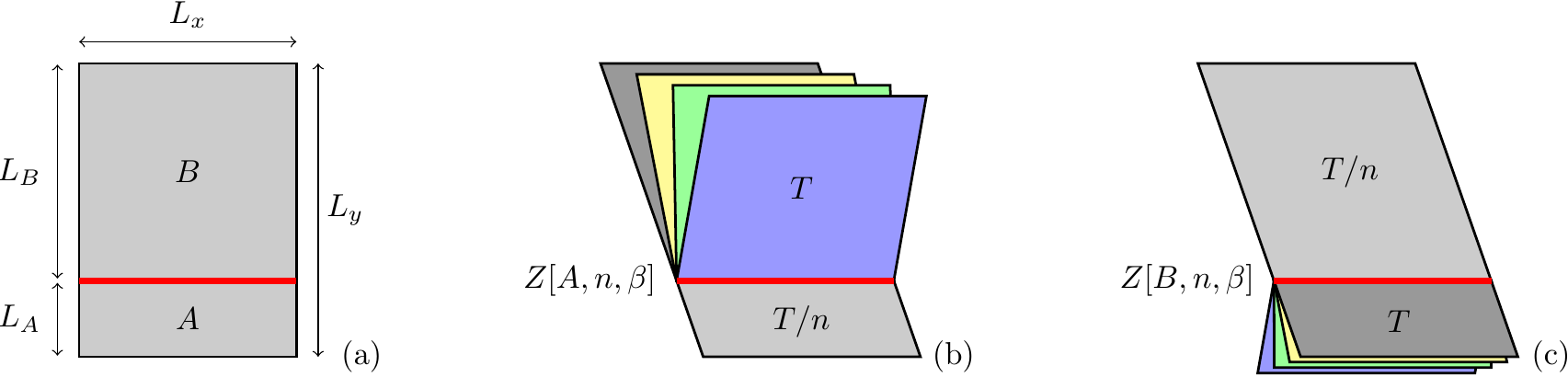}
\caption{(a) Example of a bipartition of the classical system at temperature $T=1/\beta$. The thick red line represents the boundary between $A$ and $B$. (b) and (c) Non trivial $n-$sheeted partition function $Z[A,n,\beta]$ and $Z[B,n,\beta]$ in the R\'enyi mutual information ($n=4$ shown). Each of the $n$ sheets is at temperature $T$, while the remaining one is at temperature $T/n$.}
\label{fig:replicas}
\end{figure*}

\smallskip
{\bf R\'enyi mutual information}.---
Shannon showed that entropies can be defined for any discrete probability distribution $\{p_i\}$ \cite{Shannon}. The R\'enyi entropy is
\begin{equation}
 S_n=\frac{1}{1-n}\log \left(\sum_i p_i^n\right),
 \label{RenyiShannon}
\end{equation}
where the R\'enyi index $n$ need not be an integer. In the quantum case, the $p_i$ label the eigenvalues of the reduced density matrix. In our classical case, the probabilities come from the Boltzmann weights $p_i=Z(\beta)^{-1} e^{-\beta E_i}$, where $Z(\beta)=\sum_i e^{-\beta E_i}$ is the partition function
and $\beta=1/T$ is the inverse temperature.

We cut a classical spin system into two parts $A$ and $B$ and label the spin configurations within each subsystem as $i_A$ and $i_B$ respectively. 
In $A$, the probability of observing the configuration $i_A$ is simply $ p_{i_A}=\sum_{i_B} p_{i_A,i_B}$. The R\'enyi entropy (\ref{RenyiShannon})  of this probability distribution quantifies the amount of information that can be accessed about system $A$, assuming complete knowledge of $B$. It is highly convenient to consider a more symmetric quantity, the R\'enyi mutual information (RMI):
\begin{equation}
 I_n=S_n(A)+S_n(B)-S_n(A\cup B)\ .
\end{equation}
It can be used for example to detect phase transitions, and extract the critical temperature accurately \cite{Melkoinf3}. 

Because the leading bulk contributions cancel, the RMI obeys a \emph{boundary law} \cite{Melkoinf3}:
\begin{equation}
 I_n(A,B)=a_n L+{\cal G}_n+o(1), 
  \label{eq:rmi_scaling}
\end{equation}
where $L$ is the length of the boundary between $A$ and $B$ (in Fig.~\ref{fig:replicas}(a), $L=L_x$). 
The most interesting piece of eq.~(\ref{eq:rmi_scaling}) is the subleading term ${\cal G}_n$, which for \emph{critical} systems depends on the geometry of regions $A$ and $B$. We dub it the geometric mutual information (GMI). We calculate it for two-dimensional critical systems exactly by combining renormalization-group arguments with boundary CFT. The result is universal, and can be used to identify the critical theory precisely.

\smallskip
{\bf The replicated partition functions}.---
For $n$ an integer larger than one, the RMI can be expressed as \cite{Melkoinf3}
\begin{equation}\label{eq:rmi_everything}
 I_n(A,B)=\frac{1}{1-n}\log \left(
 \frac{Z[A,n,\beta] Z[B,n,\beta]}{Z(\beta)^n Z(n\beta)}
 \right),
\end{equation}
where
\begin{equation}
\label{eq:replica_s}
 Z[A,n,\beta]=\sum_{i_A}\sum_{i_{B_1},\ldots,i_{B_n}}e^{-\beta\sum_{k=1}^n E_{i_A,i_{B_k}}}.
\end{equation}
The sum runs over all spin configurations of $n$ independent copies of the subsystem $B$, each at temperature $T$. When the interactions are local, $Z[A,n,\beta]$ can be interpreted as the partition function of  a replicated system, as is shown in Fig.~\ref{fig:replicas}(b). 
All these replicas interact with a \emph{single} copy of $A$. However in $A$, the energy is that of a system at inverse temperature $\beta^\prime=n\beta$ (and therefore temperature $T^\prime=T/n$). The analogous replicated partition function $Z[B,n,\beta]$ is shown in Fig.~\ref{fig:replicas}(c). 

The quantity $Z[A,n,\beta]$ in (\ref{eq:replica_s}) can be expressed in a form not requiring $n$ be an integer. Here we assume nearest-neighbor interactions, but the following can be generalized. We denote by {\em boundary} sites those in $A$ with neighbors in $B$; these are the sites on the thick red line in Fig.~\ref{fig:replicas}(a).  The spin configuration on the boundary sites is labeled by \mbox{$\sigma=(\sigma_1,\sigma_2,\ldots,\sigma_{L_x})$}, and we have  
\begin{equation}\label{eq:rmi_general}
 Z[A,n,\beta]=\sum_{\sigma} Z_{A}^\sigma(n\beta)\left[Z_{B}^{\sigma}(\beta)\right]^n.
\end{equation}
Here $Z_{A}^\sigma(n\beta)$ is the partition function of subsystem $A$ at temperature $T/n$, where the boundary spins are fixed to the particular configuration $\sigma$. $Z_B^\sigma(\beta)$ is the partition function of the subsystem $B$ at temperature $T$, to which the boundary sites with configuration $\sigma$ have been added. 


\comment{
First, each term has the dimension of a free energy $F=-\log Z(\beta)$. In general, it is expected to scale as 
$ F(\beta)=f_2(\beta) L^2+f_1(\beta) L+f_0(\beta)+o(1) $
where $L$ is a typical length scale in the system. $f_2(\beta)$ is the bulk free energy, while $f_1(\beta)$ is the line free energy. At the critical point $\beta=\beta_c$, the subleading term may become universal. 
}

\smallskip
{\bf Shape dependence at the critical points}.---
Our results apply to a classical lattice model with a critical point at $T=T_c$ separating two non-critical phases, one ordered and one disordered. The RMI exhibits critical behavior at both $T=T_c$ and $T=nT_c$. We first focus on $T=T_c$ in the Ising model. In the replicated picture for $Z[A,n,\beta_c]$, the subsystem $A$ is at temperature $T_c/n$ so that for $n>1$ it is in the ordered phase.
The partition function is dominated by configurations near the two ordered ones, and in the large-distance limit, one can effectively take all spins to be the same. The $n$ copies of $B$ are at temperature $T_c$ and do fluctuate. However, the different copies are coupled only via the boundary spins, which belong to system $A$ and so are all aligned here. Thus the copies do not interact, giving
\begin{equation}\label{eq:rmi_rg}
 Z[A,n,\beta_c]\approx \left[Z_B^{(+)}(\beta_c)\right]^n+\left[Z_B^{(-)}(\beta_c)\right]^n.
\end{equation}
Here $Z_B^{(+)}(\beta_c)=Z_B^{(-)}(\beta_c)$ is the partition function of system $B$ at the critical temperature with all boundary spins fixed to $+$ or $-$; the $\mathbb{Z}_2$ symmetry of the Ising model means the two are identical. 

In the continuum limit the (lattice) boundary conditions renormalize to a conformally invariant boundary condition \cite{Cardybcc}. We denote the universal parts of $Z_{A,B}^{(\pm)}(\beta_c)$ by $\mathcal{Z}_{A,B}^{\rm fix}$ and that of the whole system by $\mathcal{Z}_{A\cup B}$, giving
\begin{equation}\label{eq:univ_beta} 
 {\cal G}_n(T_c)=\frac{1}{1-n}\log \left(d\times \left[\frac{\mathcal{Z}_A^{\rm fix} \mathcal{Z}_B^{\rm fix}}{\mathcal{Z}_{A\cup B}}\right]^n\right)\ .
\end{equation}
Because of the form of (\ref{eq:rmi_everything}), the leading non-universal bulk contributions to the RMI cancel.  This applies generally when
$d$ is the number of ordered configurations in a symmetry-broken phase, e.g.\ $d=2$ and $3$ for Ising and 3-state Potts models respectively. 

The subsystem $A$ in $Z[A,n,\beta_c/n]$ is also at the critical temperature. In this case it is coupled to $n$ disordered systems, so that the conformal boundary condition on $A$ is free. 
Thus at $T=nT_c$, the GMI exhibits critical behavior, being
\begin{equation}\label{eq:univ_nbeta}
 {\cal G}_n(nT_c)=\frac{1}{1-n}\log\left(\frac{\mathcal{Z}_A^{\rm free} \mathcal{Z}_B^{\rm free}}{\mathcal{Z}_{A\cup B}}\right).
\end{equation}
Owing to (\ref{eq:rmi_general}), they also hold away from integer $n$, provided $n>1$ \footnote{Indeed the sum (\ref{eq:rmi_general}) is dominated by spin configurations with the highest magnetization, as $Z_A(n\beta_c)$ is in the ordered phase for any $n>1$.}.

Since the boundary law term in (\ref{eq:rmi_scaling}) does not depend on $L_A$ or $L_B$, the shape dependence at a critical point is given at the leading order by ${\cal G}$. This can be used to check the value of the critical temperature $T_c$ in case it is not known, as an alternative to the method of Ref.\ \onlinecite{Melkoinf3}.

\smallskip
{\bf Extracting the central charge}.---
Eqs.~(\ref{eq:univ_beta}) and (\ref{eq:univ_nbeta}) are true in any geometry in any dimension. We here focus on two dimensions, where exact expressions for the partition functions can be found using CFT.  The explicit expressions involve not only the central charge $c$ of the underlying CFT, but also $h$, the dimension of the operator that changes the boundary conditions from those on the external boundary of $A\cup B$ to those along the cut \cite{Cardybcc}.  The nicest formulas here occur for a rectangle split into two rectangles, where there are at most two places where the boundary condition changes.
In this case, the partition function for an $L_x\times L_y$ rectangle is known for all CFTs \cite{KlebanVassileva1,KlebanVassileva2,Rectanglecft2}:
\begin{equation}\label{eq:rectangle_Z}
 \mathcal{Z}=L_x^{c/4-4h} \left[f(L_y/L_x)\right]^{16h-c/2}\left[f(2L_y/L_x)\right]^{-8h}
\end{equation}
where $f$ is defined as
\begin{equation}\label{eq:function}
 f(u)=e^{-\pi u/12}\prod_{k=1}^{\infty}\left(1-e^{-2\pi k u}\right).
\end{equation}
This is related to the standard Dedekind eta function \cite{Abramowitz}, through $\eta(\tau)=f(-i \tau)$.

When the external boundary conditions are free, at $T=nT_c$ the boundary conditions along the cut are the same, and the shape function is only determined by the central charge.
Plugging (\ref{eq:rectangle_Z}) into (\ref{eq:univ_nbeta}) gives the result (\ref{eq:ad}) advertised in the introduction. 
At $T=T_c$, (\ref{eq:univ_beta}) yields
\begin{equation}\label{eq:rect_tc}
 {\cal G}_n(T_c)=\frac{n}{n-1}\left[J^{(c)}-J^{(h)}\right]\ ,
\end{equation}
where the ``central charge'' part is given by
\begin{eqnarray}\label{eq:centralchargepart}
J^{(c)}=\frac{c}{2}\log \left(\frac{f(L_A/L_x)f(L_B/L_x)}{\sqrt{L_x}f(L_y/L_x)}\right) \ ,
\end{eqnarray}
while the part proportional to the dimension of the boundary-condition changing operator is
\begin{equation}\label{bccpart}
 J^{(h)}=8h\log \left(\frac{f(L_A/L_x)^2f(L_B/L_x)^2}{L_xf(2L_A/L_x)f(2L_B/L_x)}\right)\ .
\end{equation}
For all statistical models with local positive Boltzmann weights $c$ and $h$ are positive, so both (\ref{eq:centralchargepart},\ref{bccpart}) are positive, and there is a competition between them in (\ref{eq:rect_tc}). 
For fixed spins at the external boundary, this yields \mbox{${\cal G}_n(T_c)=\frac{n}{n-1}J^{(c)}$}, and \mbox{${\cal G}_n(nT_c)=\frac{1}{n-1}[J^{(c)}-J^{(h)}]$}.

As a consequence of sharp corners in our geometry \cite{CardyPeschel}, each shape function contains an additional divergent term $\propto \log L_x$. A similar logarithm has also been identified in the RMI of certain particular 2d wave functions \cite{FradkinMoore,Zaletel,SMP3,FradkinBook}, or at finite temperature \cite{SHKM}.

\comment{\begin{equation}\label{eq:rect_2tc}
 I_n(nT_c)=\frac{J^{(c)}}{n-1}.
\end{equation}
The most natural is to let all spins fluctuate freely at the external boundary. 
; we will come back to this point later. In practice, it is easy to set the right external boundary conditions and get rid of $h$. 
}

\smallskip
{\bf Monte Carlo simulations}.---
We demonstrate the utility of the GMI by analyzing several classical critical points in Monte Carlo simulations. We use a novel method that does not require thermodynamic integration, a \emph{transfer-matrix ratio trick}. From eq.~\eqref{eq:replica_s}, we write
\begin{align}
\frac{Z[A,n,\beta]}{Z(\beta)^n} = \prod_{i=0}^{N-1} \frac{Z[A_{i+1},n,\beta]}{Z[A_{i},n,\beta]} \label{eq:ratio_trick}
\end{align}
where $A_0$ is the empty region so that $Z[A_0,n,\beta] = Z(\beta)^n$, $A_N=A$, and the $A_i$ interpolate between the two. 
Such terms have been calculated previously using methods known as ``ratio tricks'' \cite{REEQMC,Tommaso}.
This is typically done by generating a valid state from partition function $Z[A_i,n,\beta]$ and looking at the weight of the same configuration as a state from the partition function $Z[A_{i+1},n,\beta]$.

In classical systems we can calculate these ratios of partition functions by using transfer matrices; a similar idea is used for calculating partition functions on graph problems~\cite{Mehdi}. 
In a two-dimensional system, we require that the systems $Z[A_{i+1},n,\beta]$ and $Z[A_i,n,\beta]$ differ only by a one-dimensional strip of spins $S_{\{i+1\}}$.
To calculate the ratios in Eq.~\eqref{eq:ratio_trick} we treat all the spins not in $S_{\{i+1\}}$ as a fixed bath and use a transfer-matrix approach to calculate $Z[A_{i+1},n,\beta]/Z[A_i,n,\beta]$ given that fixed bath.
Since $S_{\{i+1\}}$ is one-dimensional, the two partition functions can be calculated quickly and efficiently. 
The element $Z[A_1,n,\beta]$ is the partition function of $n$ non-interacting strips each connected to their own fixed bath, while $Z[A_{i+1},n,\beta]$ is represented as (effectively) one strip of spins connected to $n$ fixed baths (one for each replica). This method gives us direct access to the shape dependence of the RMI, at any fixed $\beta$.

Since the bath of spins is fixed for the calculation, we must average this estimate of the ratio over many instances of the bath configurations, which are generated using a normal Monte Carlo (MC) procedure \cite{Mehdi}.
Each ratio is an independent calculation, thus parallelization is trivial, allowing the study of very large systems.
Restricting $S_{\{i+1\}}$ to 1d strips ensures 
that the largest amount of time is spent generating random states.
In a typical MC procedure this scales as $\mathcal{O}(N)$, where $N$ is the number of spins in the system.
This implies that the calculation of $S_n(A)$ should scale as $\mathcal{O}(n N L_A)$,

\smallskip
{\bf Comparing numerics to theory}.---
To compare numerical simulations to the theoretical results (\ref{eq:ad},\ref{eq:rect_tc}), we consider discrete spin systems on a $L\times L$ square lattice, cut into two rectangles of respective sizes $\ell\times L$ and $(L-\ell)\times L$. The spins are not constrained at the external boundary and so fluctuate freely. 
We compute the second RMI $I_2$ at $T=T_c$ and $T=2T_c$, using the MC techniques described above.
The shape dependence is completely determined by the geometry; the central charge is the \emph{leading} coefficient, as opposed to its appearance as a subleading term in the free energy \cite{Cardyc,Affleckc}. Our method has other nice features: it does not require studying off-critical behavior as in Refs.\ \onlinecite{cMC1,CornerPotts}, nor implementing a lattice version of the stress-tensor \cite{cMC2}.

\begin{figure}[tbp]
  \includegraphics[width=8cm]{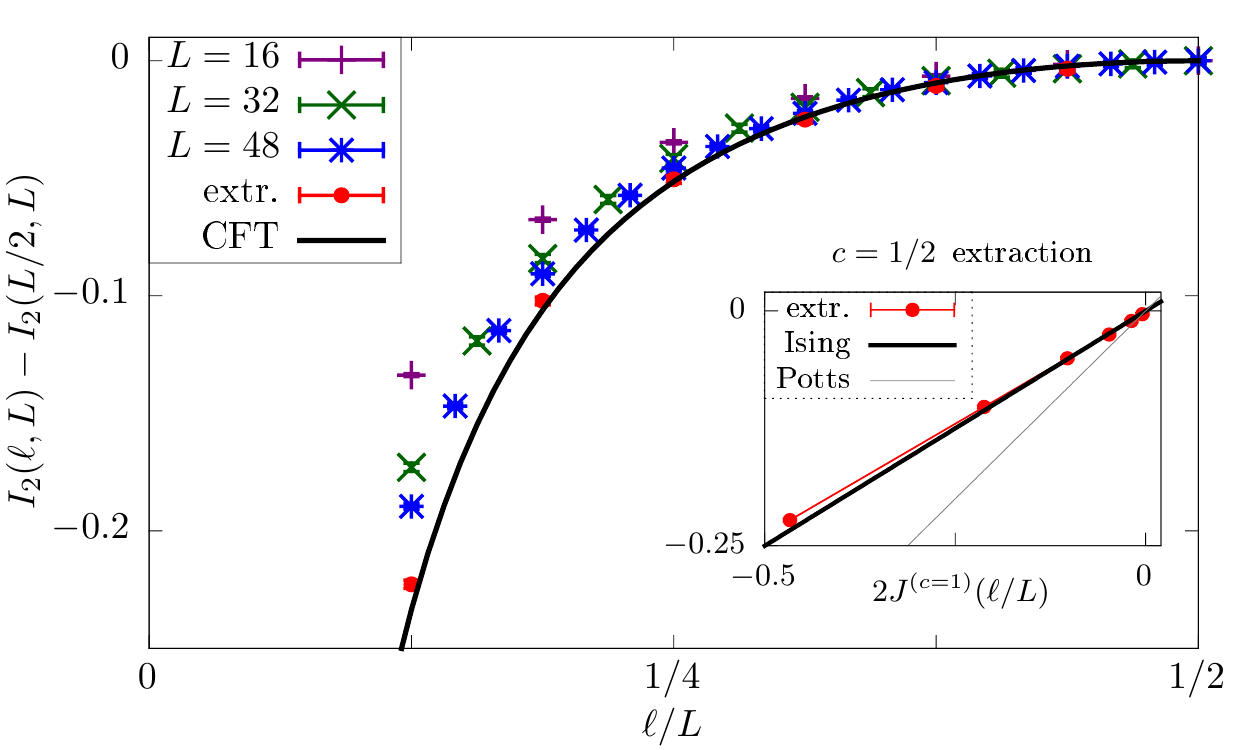}
  \caption{Second RMI for the Ising model at $T=2T_c$, for system sizes $L=16,32,48$, and comparison with CFT (black curve). Red circles represent the extrapolated data discussed in the text. 
  Inset: Ising central charge extraction. The data can be seen to agree very well with $c=1/2$. We also show the line corresponding to the $Q=3$ Potts model for comparison.
 }
  \label{fig:Ising_2tc}
 \end{figure}
 \begin{figure}[bp]
  \includegraphics[width=8cm]{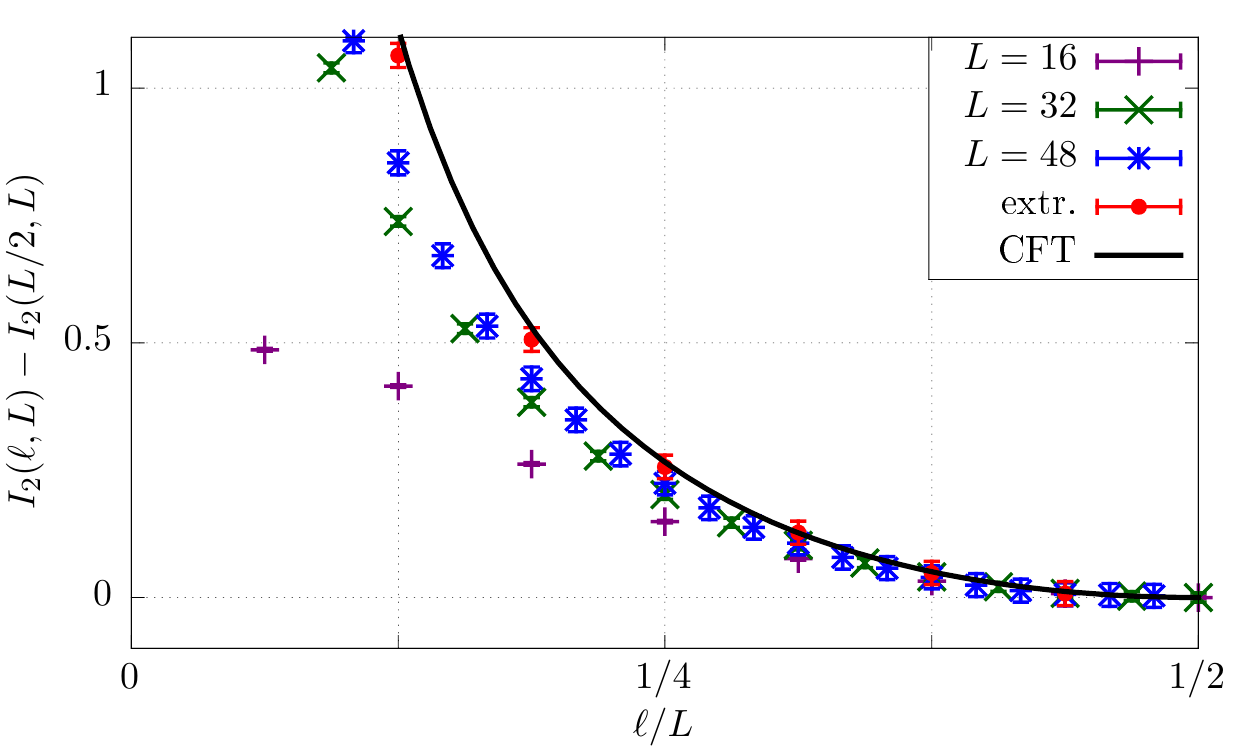}
  \caption{Second RMI for the Ising model at $T=T_c$, for system sizes $L=16,32,48$, extrapolation, and comparison with CFT (black curve). Due to the presence of boundary-condition changing operators, the curve is  flipped compared to Fig.~\ref{fig:Ising_2tc}.}
  \label{fig:Ising_tc}
 \end{figure}

 We first focus on the Ising model. At $T=2T_c$, the RMI is given by Eq.~(\ref{eq:ad}), and so gives direct access to the central charge, $c=1/2$ here. We plot $I_2(\ell,L)-I_2(L/2,L)$ in Fig.~\ref{fig:Ising_2tc}. At small aspect ratio $\ell/L$ finite-size effects should be large, so we extrapolate by fitting the data to $a+bL^{-1}$ for each $\ell/L$. The presence of this $L^{-1}$ correction is expected in geometries with sharp corners (see e.g.~\cite{Zaletel,Isingrectanglenum1,logloverl}).  The agreement with the shape dependence of the CFT with $c=1/2$  is excellent. 
One way to extract the central charge without knowing it {\it a priori} is to plot the extrapolated data as a function of the $c=1$ CFT shape. One then expects to see a straight line with slope $c$. As can be seen in the inset, the agreement with $c=1/2$ remains excellent, our best estimate being $c=0.49(1)$. 

\begin{figure}[tbp]
  \includegraphics[width=8cm]{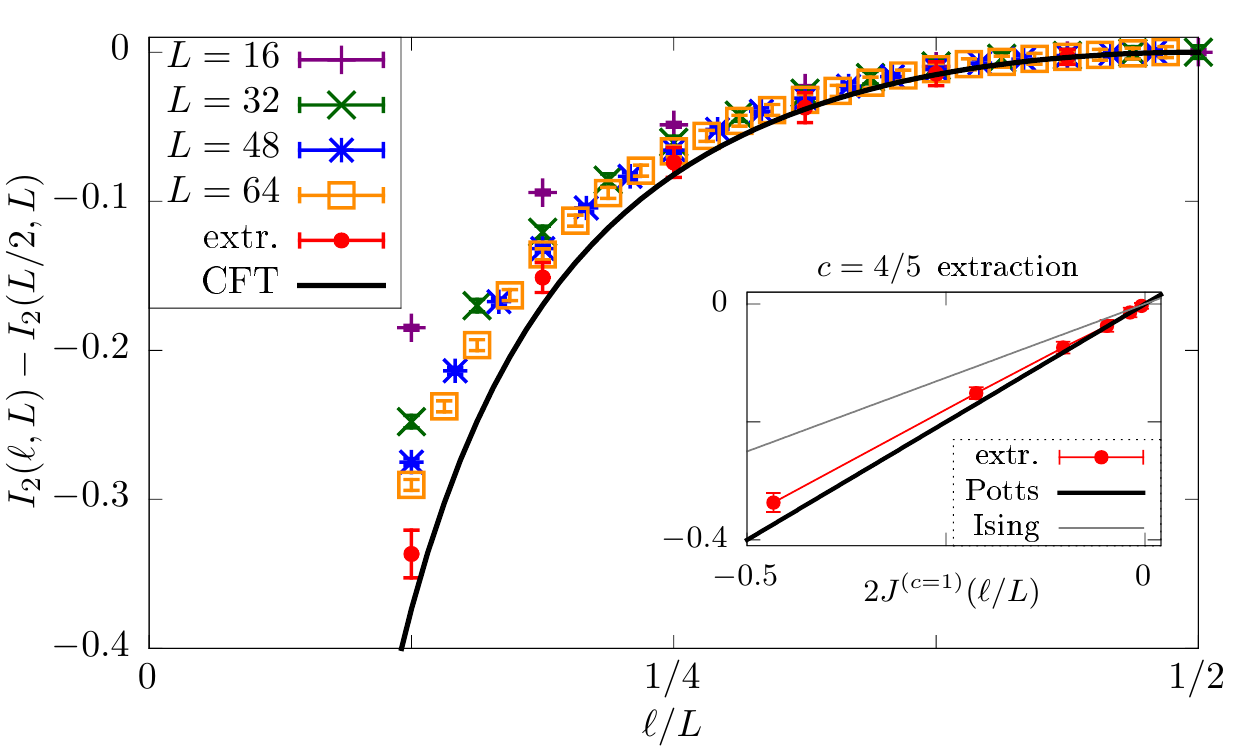}
  \caption{Second RMI for the $Q=3$ Potts model at $T=2T_c$, for system sizes $L=16,32,48,64$, and comparison with CFT (black curve). Red circles are the extrapolated data. Inset: central charge extraction. The data is compatible with the expected $c=4/5$ in the thermodynamic limit. The Ising slope is also shown for comparison.}
  \label{fig:Potts_2tc}
 \end{figure}

Our numerical results for $T=T_c$ are shown in Fig.~\ref{fig:Ising_tc}.
Here the boundary conditions along the cut are fixed, so with external free boundary conditions, Eq.~(\ref{eq:rect_tc}) applies.  The conformal dimension of the operator that changes between fixed and free boundary conditions in the Ising model is $h=1/16$ \cite{Cardybcc}, giving a curve in agreement with our numerics.  Interestingly, the effect of a non-zero $h$ counterbalances the central charge part, and is even enough to flip the shape, akin to the even/odd effect in 2d RVB wave functions \cite{us}.

To illustrate the generality of our method, we checked our results in the three-state Potts model, a strongly interacting model not mappable to free fermions like Ising.  The numerical results at $T=2T_c$ are shown in Fig.~\ref{fig:Potts_2tc}. These agree well with  Eq.~(\ref{eq:ad}) using the three-state Potts central charge $c=4/5$. Given that finite-size effects in $Q$-state Potts  increase with $Q$ \cite{clusterPotts}, our fit of $c=0.75(5)$ is good. We also checked numerically that at $T=T_c$, Eq.~(\ref{eq:rect_tc}) holds with the correct boundary exponent $h=1/8$ \cite{Cardybcc}; similarly to Ising the shape is also flipped.

\smallskip
{\bf Conclusion}.---
We have defined the geometric mutual information and computed it for 2d critical classical systems. 
The GMI is easy to measure in classical simulations using the transfer-matrix ratio trick, and so can be used to identify the universality class.
The expressions eq.\ \ref{eq:univ_beta},\ref{eq:univ_nbeta} relating the GMI to partition functions hold in any geometry in any dimension, and we have verified numerically the exact results for the cylinder and torus as well as the rectangle described above.

An interesting future direction is to compute the RMI for the $XY$ model, which never orders at low temperature. Here the replicated picture yields a non trivial gluing of $c=1$ compact-boson CFTs with different radii. It also would be illuminating to study further the close connection
of the GMI with the entanglement entropy of certain 2d quantum wave functions, for example the transitions that have been observed as a function of the R\'enyi index \cite{SMP3,Zaletel,Sondhi}. It could shed light on a vexing problem occurring in the singular limit $n\to 1$, where universal subleading terms \cite{SFMP,AletShannon} exhibit mysterious boundary critical behavior in infinite geometries in the Ising case \cite{SMP2,Zaletel}.

{\it Acknowledgments}.--- We thank R. Bondesan, J. Cardy and E. Fradkin for valuable insights.  
We especially thank Mehdi Molkaraie for drawing our attention to the methods of Ref.~\cite{Mehdi}.
The simulations were performed on the computing facilities of SHARCNET.  JMS and PF are supported by National Science Foundation grant DMR/MPS1006549.  RGM acknowledges support from NSERC, the Canada Research Chair program, the John Templeton Foundation, and the Perimeter Institute (PI) for Theoretical Physics. Research at PI is supported by the Government of Canada through Industry Canada and by the Province of Ontario through the Ministry of Economic Development \& Innovation.

\bibliography{transfer2}

\end{document}